\documentclass[10pt,onecolumn]{article}

\newcommand{\Lm}{\mbox{\L}}
 
\renewcommand{\title}[1]{\null\vspace{25mm}\noindent{\Large{\bf #1}}\vspace{10mm}}
\newcommand{\authors}[1]{\noindent{\large #1}\vspace{20mm}}
\newcommand{\address}[1]{{\center{\noindent #1\vspace{10mm}}}}
\renewcommand{\abstract}[1]{\vspace{17mm}
\noindent{\small{\em Abstract.} #1}\vspace{2mm}}     
 
\newcommand{\be}{\begin{equation}}
\newcommand{\ee}{\end{equation}}
\newcommand{\ba}{\begin{array}}
\newcommand{\ea}{\end{array}}
\newcommand{\bea}{\begin{eqnarray}}
\newcommand{\eea}{\end{eqnarray}}

\newcommand{\nm}{\nonumber}

\newcommand{\Li}{{\cal L}_\tau}
\newcommand{\del}{\delta_\tau}
\newcommand{\id}{i_\tau}
\newcommand{\g}{g(\tau)}
\newcommand{\tr}{{\rm tr}}
\newcommand{\trbr}[1]{\tr \left(#1\right)}
\newcommand{\intM}[1]{\int_{{\cal M}_#1}}

\newcommand{\G}[1]{\Gamma^{(#1)}}

\newcommand{\ie}{{\it i.\,e.\ }}
\newcommand{\eg}{{\it e.\,g.\ }}
\newcommand{\cf}{{\it c.\,f.\ }}
\newcommand{\cu}[1]{{}^{#1}\!}

\newcommand{\pui}{{p_i}}
\newcommand{\dpui}{{d-\pui-1}}
\newcommand{\rui}{{r_i}}

\newcommand{\BRSTs}{BRST-trans\-for\-mations }
\newcommand{\VSUSY}{VSUSY-trans\-for\-mation }
\newcommand{\VSUSYs}{VSUSY-trans\-for\-mations }

\newcommand{\W}[2]{\frac{\delta #1}{\delta #2}}

\newcommand{\bh}{\frac{1}{2}}

\newcounter{saveeqn}

\newcommand{\captionit}[1]{\caption{\small\sl{#1}}}

\begin{document}   \setcounter{table}{0}
 
\begin{center}
\hspace*{\fill}{{\normalsize \begin{tabular}{l}
                              {\sf hep-th/0004137}\\
			      {\sf REF. TUW 00-12}\\
			      {\sf \today}
                              \end{tabular}   }}

\title{Symmetry content of a generalized $p$-form model \\ of Schwarz-type in $d$ dimensions}
\renewcommand{\thefootnote}{\fnsymbol{footnote}}

\addtocounter{footnote}{1}
\authors {Thomas Pisar\footnote{Work supported by the "Fonds zur F\"orderung der Wissenschaflichen Forschung", under Project Grant Number P11582-PHY.}}
\vspace{-20mm}
\renewcommand{\thefootnote}{\arabic{footnote}}       
\addtocounter{footnote}{-2}

\address{Institut f\"ur Theoretische Physik, TU Wien\\
      Wiedner Hauptstra\ss e 8-10, A-1040 Wien, Austria}
\end{center} 
\thispagestyle{empty}
\vspace{-20mm}
\abstract{We derive the vector supersymmetry and the \L-symmetry transformations for the fields of a generalized topological $p$-form model of Schwarz-type in $d$ space-time dimensions.}


\section{General setup}

BF models and their properties are a widely investigated area in the literature \cite{BF}. In particular in ref.~\cite{Lucchesi:PiguetSorella,PiguetSorella} BF models in arbitrary dimensions are considered in great detail. In ref.~\cite{Wallet:Ikemori:Dayi} a conclusive formalism in order to derive the minimal action in the framework of Batalin and Vilkovisky (BV) is presented. Based on these concepts ref.~\cite{Baulieu} introduces a possible generalization to more generic models of Schwarz-type. In this paper we present the BRST-variations, the vector supersymmetry (VSUSY) as well as a scalar supersymmetry, denoted as \L-symmetry, of the proposed model. \newline
The classical, invariant action of such a generalized $p$-form model in $d$ space-time dimensions is give by\footnote{In the following we omit the wedge product sign $\wedge$.}
\bea
	S_{inv}=\intM{d}\left\{B_pF_2+\sum_iX_{p_i}DY_{\rui}\right\},
\label{invaction}
\eea
where $p=d-2$, $\rui=\dpui$ and $\sum_i$ denotes the summation over an arbitrary set of pairs of $\pui$- and $\rui$-form fields. $F_2=dA+AA$ is the usual curvature of the connection one-form $A$ and the covariant derivative is $D=d+[A,.]$. All commutators $[.,.]$ are understood in a graded sense. The fields $B_p$, $X_\pui$ and $Y_\rui$ exhibit more or less reducible gauge symmetries, \eg $\delta_{\lambda_{p-1}} B_p=D\lambda_{p-1}$. The algebra of gauge-transformations closes on-shell. The gauge-invariant equations of motion imply the zero-curvature conditions. \newline
Following the guideline of \cite{Baulieu} we define pairs of generalized forms which are called dual to each other\footnote{Upper indices label the ghost-number and lower ones denote the form-degree.}
\be\ba{rclrcl}
	\tilde B_p&=& B_d^{-2}+B_{d-1}^{-1}+B_p+\ldots+B^p, &\tilde A& =& A_d^{1-d}+\ldots+A_2^{-1}+A+c,\nm \\
	\tilde X_\pui&=& X_d^{\pui-d}+\ldots+X_\pui+\ldots+X^\pui, &\tilde Y_\rui& =& Y_d^{\rui-d}+\ldots+Y_\rui+\ldots+Y^\rui.
\label{genexp}
\ea\ee
For later convenience we cast all fields of \eg $\tilde X_\pui$ with negative Faddeev-Popov ($\Phi\Pi$) charge into $\check X_\pui$, whereas the contributions with positive or zero $\Phi\Pi$-charge are collected in $\hat X_\pui$. Hence, a total expansion is $\tilde X_\pui=\check X_\pui+ \hat X_\pui$. The denotation of dual fields becomes more evident if one observes that the fields with negative $\Phi\Pi$-charge, serve as antifields in the sense of Batalin and Vilkovisky for the elements of the dual partner with positive ghost-degree, \ie $\check X_\pui=\pm (\hat Y_\rui)^*$, and vice versa\footnote{The superscript ${}^*$ denotes antifields.}. The classical gauge and ghost fields can be addressed by $\Phi^a$, whereas the corresponding antifields are collected in $\Phi^*_a$.

\section{BRST-symmetry and BV-action}

With the definition of a generalized exterior derivative $\tilde d=d+s$, with $s$ denoting the BRST-differential, we can define a generalized covariant derivative $\tilde D=\tilde d+[\tilde A,.]$ and construct the curvatures
\bea
	\tilde F_2=\tilde d \tilde A+\tilde A\tilde A, \quad	\tilde G_{p+1} = \tilde D \tilde B_p, \quad	\tilde H_{\pui+1} = \tilde D \tilde X_\pui, \quad	\tilde I_{\rui+1} = \tilde D \tilde Y_\rui.
\label{curvature}
\eea
The \BRSTs of the fields are determined through so-called horizontality conditions \cite{Thierry-Mieg:1980aa} which read
\be\ba{llll}
	\tilde F_2=0, & \tilde G_{p+1}=\sum_i(-1)^\pui [\tilde X_\pui,\tilde Y_\rui],& \tilde H_{\pui+1}=0,& \tilde I_{\rui+1}=0.
\ea\ee
With the definitions $F_2^{\tilde A}=d\tilde A+\tilde A\tilde A$ and $D^{\tilde A}=d+[\tilde A,.]$ this yields
\be\ba{rclrcl}
	s \tilde A &=& -F_2^{\tilde A}, & s \tilde B_p &=& -D^{\tilde A}\tilde B_p+\sum_i(-1)^\pui [\tilde X_\pui,\tilde Y_\rui], \nm \\
	s \tilde X_\pui &=& -D^{\tilde A}\tilde X_\pui, & s \tilde Y_\rui &=& -D^{\tilde A}\tilde Y_\rui,
\label{BRST}
\ea\ee
The above \BRSTs admit the cocycle equation 
\bea
	\tilde d \; \trbr{\tilde B_pF_2^{\tilde A}+\sum_i\tilde X_\pui D^{\tilde A}\tilde Y_\rui}=0. 
\eea
This leads to the BRST-invariant, minimal BV-action
\bea
	S_{min}=\intM{d}\left.\left\{\tilde B_p F_2^{\tilde A}+\sum_i \tilde X_\pui D^{\tilde A} \tilde Y_\rui\right\}\right|_d^0.
\label{actionmin}
\eea
An expansion in the generic fields yields
\bea
	S_{min} &=& S_{inv}+\intM{d}\left\{\check B_p(-s\hat A)+\check A (-s\hat B_p)+\sum_i\left(\check X_\pui (-s\hat Y_\rui)+(-1)^{d(\pui+1)}\check Y_\rui(-s\hat X_\pui) \right)\right. \nm\\
		&+& \left.\left. \sum_i(-1)^\pui\check A\left([\hat X_\pui,\check Y_\rui]+[\check X_\pui,\hat Y_\rui]\right)-\bh\hat B_p[\check A,\check A]\right\}\right|_d^0. 
\label{minaction}
\eea
The latter action induces the antifield identification
\bea
 	\check B_p=(\hat A)^*,\quad  \check A = (\hat B_p)^*, \quad \check X_\pui = (\hat Y_\rui)^*, \quad  \check Y_\rui = (-1)^{d(\pui+1)}(\hat X_\pui)^*,
\eea
in order to ensure coincidence with the \BRSTs obtained from 
\bea
	s\Phi^*_a=-\W{S_{min}}{\Phi^a}, \qquad s\Phi^a=-\W{S_{min}}{\Phi^*_a}.
\eea

\section{Gauge-fixing}

Similarly to the BF model we need to introduce for each classical gauge field a BV-pyramid (\cf Table \ref{pyramidX} and \ref{pyramidY}). From the gauge-fixing point of view the dual fields $B_p$ and $A$ can be considered as an ordinary $p$- and one-form field. Hence, we define $\tilde X_{d-2}\equiv \tilde B_p$ and $\tilde Y_1 \equiv \tilde A$ and according definitions for the antighost and multiplier fields. Moreover, for the sake of a compact notation when giving the gauge-fermion, we let the antighost fields to the lowest order $n$ be the classical gauge and ghost fields, \ie $\bar v^q_{\pui-q}\equiv X^q_{\pui-q}$ and $\bar w^q_{\rui-q}\equiv Y^q_{\rui-q}$. The gauge-fixing fermion $\Psi_{gf}$ then looks like
\begin{table}[t]
\setlength{\unitlength}{1cm}
\begin{center}
\begin{minipage}[t]{8.4cm}
\begin{center}
\begin{tabular}{ccccccccc} 
      	&&& $X_\pui$& \\
      	&&$\bar v_{\pui-1}^{-1}$&& $X_{\pui-1}^1$ & \\
	&$\bar v_{\pui-2}^{-2}$&&$\bar v_{\pui-2}$&&$X_{\pui-2}^2$  \\
	\ldots &&&&&& \ldots
\end{tabular}
\captionit{$X_\pui$ pyramid}
\label{pyramidX}
\end{center}
\end{minipage}
\hfill
\begin{minipage}[t]{8.4cm}
\begin{center}
\begin{tabular}{ccccccccc} 
        &&& $Y_\rui$& \\
        &&$\bar w_{\rui-1}^{-1}$&& $Y_{\rui-1}^1$ & \\
	&$\bar w_{\rui-2}^{-2}$&&$\bar w_{\rui-2}$&&$Y_{\rui-2}^2$  \\
	\ldots &&&&&& \ldots
\end{tabular}
\captionit{$Y_\rui$ pyramid}
\label{pyramidY}
\end{center}
\end{minipage}
\end{center}
\end{table}
\bea
	\Psi_{gf} &=& \intM{d}\left\{\sum_i\sum_{n=1}^\pui\sum_{q=0}^{\pui-n}\bar v_{\pui-q-n}^{\gamma(n)}\left(\cu{n}\alpha_{(q)}*\kappa_{\pui-q-n}^{\gamma(n+1)}+d*\bar v_{\pui-q-n+1}^{\gamma(n+1)}\right)\right. \nm \\
		  &+& \left. \sum_i\sum_{n=1}^\rui\sum_{q=0}^{\rui-n}\bar w_{\rui-q-n}^{\gamma(n)}\left(\cu{n}\beta_{(q)}*\lambda_{\rui-q-n}^{\gamma(n+1)}+d*\bar w_{\rui-q-n+1}^{\gamma(n+1)}\right)\right\}, 
\eea
where ${}*$ is the Hodge-star operator, $\cu{n}\alpha_{(q)}, \cu{n}\beta_{(q)} \in {\bf R}$ are some arbitrary parameters and $\gamma(n=2k)=q$ or $\gamma(n=2k+1)=-q-1$. The implementation of external sources $\rho_a^*$ implies a further contribution to the gauge-fermion
\bea
	\Psi_{ext}	&=& \intM{d}\sum_i(-1)^{(d+1)\pui}\left(\sum_{q=0}^\pui (-1)^{d+1}X^q_{\pui-q}\tau_{d-\pui+q}^{*-q-1}+\sum_{q=0}^\rui Y^q_{\rui-q}\eta_{d-\rui+q}^{*-q-1}\right). \nm
\eea
The total gauge-fermion is $\Psi=\Psi_{gf}+\Psi_{ext}$. The corresponding multiplier fields to the antighosts $\bar v_p^q$ and $\bar w_p^q$ are $\kappa_p^{q+1}$ and $\lambda_p^{q+1}$ respectively. The auxiliary contribution is given by
\bea
	S_{aux} &=& -\intM{d}\sum_i\left(\sum_{n=1}^\pui\sum_{q=0}^{\pui-n}\left(\bar v_{\pui-q-n}^{\gamma(n)}\right)^*\kappa_{\pui-q-n}^{\gamma(n)+1}+\sum_{n=1}^\rui\sum_{q=0}^{\rui-n}\left(\bar w_{\rui-q-n}^{\gamma(n)}\right)^*\lambda_{\rui-q-n}^{\gamma(n)+1}\right). \nm
\eea
The BRST-doublet fields are collected in $(\bar C^\alpha,\Pi^\alpha)$ and together with the fields of $\Phi^a$ they may by addressed by $\Phi^A$. The non-minimal solution of the BV-masters equation is given by $S_{nm}=S_{min}+S_{aux}$. The antifields can be expressed as functionals of the fields via the equation 
\bea
	\Phi^*_A[\Phi^A]=(-1)^{(d+1)|\Phi^A|+d}\W{\Psi}{\Phi^A}.
\label{exanti}
\eea
This admits the elimination of the antifields of the action in order to get
\bea
	\G{0}=\left.S_{nm}\right|_{\Phi^*_A[\Phi^A]}=S_{inv}+S_{gf}+S_{ext},
\eea
where $S_{gf}=s\Psi_{gf}+S^{mod}_{gf}$ and $S_{ext}=s\Psi_{ext}+S_{ext}^{mod}$. The additional contribution $S^{mod}=S^{mod}_{gf}+S_{ext}^{mod}$ appears since the \BRSTs of the fields (\ref{BRST}) exhibit an antifield-dependency. The structure of $S^{mod}$ can be seen from the last line in (\ref{minaction}). Although, $S^{mod}$ can not be written as a BRST-exact expression, it however, does not spoil the topological character of the model, due to its metric independence. With equation (\ref{exanti}) the antifields can also be eliminated in the generalized field expansions (\ref{genexp}). 

\section{Vector supersymmetry}

\subsection{Derivation}

The above concepts provide a neat formalism to derive the VSUSY \cite{Gieres:2000pv}, $\del=\tau^\mu\delta^{-1}_\mu$, where $\tau$ is a constant, BRST-invariant, even graded vector-field\footnote{Henceforth ${\cal M}_d$ is considered as a flat space-time manifold.}. The \VSUSYs satisfy the algebra
\bea
	[s,\del]=\Li=[d,\id],
\label{algebra}
\eea 
where $\Li$ is the Lie-derivative and $\id$ is the interior product along $\tau$. The algebra (\ref{algebra}) suggests an equivalence between $\id$ and $\del$, yielding for the $\del$-transformations
\bea
	\del \tilde X_\pui=\id\tilde X_\pui,\quad\del\tilde Y_\rui=\id\tilde Y_\rui.
\label{susyfield}
\eea
This determines the \VSUSYs of the classical gauge and ghost fields but also of the antighost fields at the first reducibility level. However, in order to describe the $\del$-transformations of the higher reducibility antighost fields we need to collect also the antighosts with positive $\Phi\Pi$-charge together with their corresponding antifields in generalized forms\footnote{The reasoning is only given for the $X_\pui$-sector, but the same arguments hold for the antighost fields $\bar w^q_{\rui-n-q}$.}
\bea
	\hat{\bar v}_{\pui-n}=\sum_{q=0}^{\pui-n}\bar v^q_{\pui-n-q},\qquad (\hat{\bar v}_{\pui-n})^*=\sum_{q=0}^{\pui-n}(\bar v^q_{\pui-n-q})^*,\qquad  n=2k. 
\label{antighostantifield}
\eea 
Although only the antighosts with positive $\Phi\Pi$-charge are cast into this scheme, the antighost fields with negative ghost-degree come into play automatically via the elimination of the antifields (\ref{exanti}). The \VSUSYs now follow from the proposed equivalence of the $\del$-operation and the interior product $\id$ in the space of generalized forms and their duals, thus we get
\bea
	\del \hat{\bar v}_{\pui-n}=\id \hat{\bar v}_{\pui-n}, \qquad \del (\hat{\bar v}_{\pui-n})^*=\id (\hat{\bar v}_{\pui-n})^*.
\label{susyanti}
\eea
This determines the \VSUSYs of the remaining antighost fields, but also the gauge-parameters $\cu{n}\alpha_{(q)}=\cu{n}\beta_{(q)}=(-1)^d$ for $n=2k$. 

\subsection{Explicit results}

The detailed results for the elements of $\Phi^a$ yield from (\ref{susyfield})
\be\ba{rcll}
	\del X_\pui	&=& -\id\left(\eta_{\pui+1}^{*-1}-(-1)^{(d+1)\pui+d}*d\bar w^{-1}_{\rui-1}\right), \\ 
	\del X^q_{\pui-q} &=& \id X^{q-1}_{\pui-q+1}, & \quad  q=1,\ldots,\pui,\\
	\del Y_{\rui} 	&=& -\id\left(\tau_{\rui+1}^{*-1}-(-1)^{d+\pui+1}* d\bar v^{-1}_{\pui-1}\right), \\
	\del Y^q_{\rui-q} &=& \id Y^{q-1}_{\rui-q+1}, & \quad  q=1,\ldots,\rui. 
\label{susy1}
\ea\ee
The antighost field transformations are determined through (\ref{susyanti})
\be\ba{rcll}
	\del \bar v_{\pui-n}&=&0, \\
	\del \bar v^q_{\pui-q-n}&=&\id \bar v^{q-1}_{\pui-q-n+1}, & \quad q=1,\ldots,\pui-n, \\
	\del \bar v^{-q-1}_{\pui-q-n} &=& (-1)^{d+\pui+q+1}\g \bar v^{-q-2}_{\pui-q-n-1}, & \quad q=0,\ldots, \pui-n-1, \\
	\del \bar v^{-\pui+n-1}_0 &=& 0, 
\ea\ee
and
\be\ba{rcll}
	\del \bar w_{\rui-n}&=&0, \\ 
	\del \bar w^q_{\rui-q-n}&=&\id \bar w^{q-1}_{\rui-q-n+1},  &\quad q=1,\ldots,\rui-n, \\ 
	\del \bar w^{-q-1}_{\rui-q-n} &=& (-1)^{d+\rui+q+1}\g \bar w^{-q-2}_{\rui-q-n-1}, & \quad q=0, \ldots, \rui-n-1, \\
	\del \bar w^{-\rui+n-1}_0 &=& 0,
\ea\ee
where the Hodge-star ${}*$ intertwines between the interior product and the one-form $\g=\tau_\mu dx^\mu$ in the way $\id*\alpha_p=(-1)^p*\g\alpha_p$. The corresponding multiplier field transformations can be obtained from the algebra (\ref{algebra}) through 
\bea
	\del \Pi^\alpha=\Li\bar C^\alpha-s\del\bar C^\alpha,
\eea
for some element of the BRST-doublets $(\bar C^\alpha, \Pi^\alpha)$. Furthermore, the $\del$-variations of the external sources are
\be\ba{rcll}
	\del \tau_{d-\pui+q}^{*-q-1}&=& \id \tau_{d-\pui+q+1}^{*-q-2}, &\quad q=0,\ldots,\pui-1, \\
	\del \tau_d^{*-\pui-1}&=&0 , \\
	\del \eta_{d-\rui+q}^{*-q-1}&=& \id \eta_{d-\rui+q+1}^{*-q-2}, &\quad q=0,\ldots,\rui-1, \\
	\del\eta_d^{*-\rui-1}&=&0.
\label{susy3}
\ea\ee
The algebra (\ref{algebra}) of the \VSUSYs of the classical gauge fields closes on-shell 
\bea
	\left[ s,\del \right] X_\pui = \Li X_\pui -(-1)^{d(\pui+1)} \id \W{\G{0}}{Y_\rui}, \qquad \left[ s,\del \right] Y_\rui = \Li Y_\rui-\id \W{\G{0}}{X_\pui}. 
\eea
In general we describe the symmetry content of a model with a Ward-operator
\bea
	{\cal W}^I=\intM{d}\sum_\varphi\delta^I\varphi\W{}{\varphi},
\label{WI}
\eea
where $\delta^I\varphi$ denotes the field-transformations under the symmetry $I$ and $\varphi$ stands for all fields characterizing the model in question. In this sense we define a Ward-operator ${\cal W}_\tau$ according to the above $\del$-transformations (\ref{susy1})--(\ref{susy3}). By choosing the remaining gauge-parameters $\cu{2k+1}\alpha_{(q)}=\cu{2k+1}\beta_{(q)}=0$, the application of ${\cal W}_\tau$ onto $\G{0}$ leads to a linear breaking term in the quantum fields 
\bea
	{\cal W}_\tau\G{0}=\Delta_\tau,
\eea
where
\bea
	\Delta_\tau	&=& \intM{d}\sum_i (-1)^\pui \left\{(-1)^{d+1}\left(\sum_{q=0}^\pui \tau_{d-\pui+q}^{*-q-1}\Li X_{\pui-q}^q+\kappa_{\pui-1}d*\id\eta_{\pui+1}^{*-1}\right)\right. \nm \\
			&+&\left.\left(\sum_{q=0}^\rui \eta_{d-\rui+q}^{*-q-1}\Li Y_{\rui-q}^q+\lambda_{\rui-1}d*\id\tau_{\rui+1}^{*-1}\right)\right\}. 
\eea

\section{\L-symmetry}

\subsection{General setup}

For the following section we assume that the action is complete in the sense, that it contains all possible types of $p$-forms that are allowed, but where no particular $p$-form shall occur more than once. The definitions $I_1^{\tilde Y_0}= D^{\tilde A}\tilde Y_0$, $I_2^{\tilde Y_1}\equiv F_2^{\tilde A}$, $I_3^{\tilde Y_2}=D^{\tilde A}\tilde Y_2$, \ldots, $I_r^{\tilde Y_{r-1}}=D^{\tilde A}\tilde Y_{r-1}$ admit to write (\ref{actionmin}) as
\bea
	S_{min}=\intM{d}\left.\left\{\sum_{i=1}^r \tilde X_{d-i} I_i^{\tilde Y_{i-1}}\right\}\right|_d^0.
\label{actionredef}
\eea
The upper limit is given by $r=d/2$ for even or $r=(d+1)/2$ for odd dimensions. We define a scalar transformation \L~with $\Phi\Pi$-charge -1 with the algebra $[\Lm,s]=0$. The \L-transformations act on the generalized fields as follows
\be\ba{rcll}
	\Lm \tilde X_{d-i}&=&\tilde X_{d-i-1}, &\quad i=1,\ldots, r-1, \\ 
	\Lm \tilde X_{d-r}&=&0,\\
	\Lm \tilde Y_0 &=& 0, & \\
	\Lm \tilde Y_{i-1}&=&(-1)^{d+i}\tilde Y_{i-2}, &\quad i=2,\ldots, r.
\ea\ee

\subsection{Explicit results}

The elimination of the antifields via (\ref{exanti}) yields the explicit transformation properties. The classical fields transform as
\be\ba{rcll}
	\Lm X_{d-i} &=& -\eta_{d-i}^{*-1}+(-1)^{(d+1)i+1}*d\bar w^{-1}_{i-1}, &\quad i=1,\ldots, r-1, \\
	\Lm X_{d-r} &=& 0, \\
	\Lm Y_0 &=& 0, \\
	\Lm Y_{i-1} &=& (-1)^{(d+1)i}\left(-\tau_{i-1}^{*-1}+(-1)^{(d+1)i+d}*d\bar v^{-1}_{d-i}\right), &\quad i=2,\ldots, r.
\label{L1}
\ea\ee
The ghosts vary under the \L-symmetry like
\be\ba{rclll}
	\Lm X_{d-i-q}^q &=& X_{d-i-q}^{q-1}, &\quad i=1,\ldots, r-1, &\quad q=1,\ldots,d-i,\\
	\Lm X_{d-r-q}^q &=& 0, &&\quad q=1,\ldots,d-r,\\
	\Lm Y_{i-1-q}^q &=& (-1)^{d+i} Y_{i-1-q}^{q-1}, &\quad i=2, \ldots, r, &\quad q=1,\ldots,i-1.
\ea\ee
The antighost fields transform as
\be\ba{rclll}
	\Lm \bar v_{d-i-n} &=& 0, &\quad i=1,\ldots,r,\\
	\Lm \bar v_{d-i-n-q}^q  &=& \bar v_{d-i-n-q}^{q-1}, &\quad i=1,\ldots,r-1, &\quad q=1,\ldots,d-i-n,\\ 
	\Lm \bar v_{d-r-n-q}^q&=&0, &&\quad q=1,\ldots,d-r-n,\\
	\Lm \bar v_{d-1-n-q}^{-q-1} &=& 0, &&\quad q=0,\ldots,d-1-n\\
	\Lm \bar v_{d-i-n-q}^{-q-1} &=& (-1)^{i+1} \bar v_{d-i-n-q}^{-q-2}, &\quad i=2,\ldots,r, &\quad q=0,\ldots d-i-n, 
\ea\ee
and 
\be\ba{rclll}
	\Lm \bar w_{i-1-n} &=& 0, &\quad i=1,\ldots,r,\\
	\Lm \bar w_{i-1-n-q}^q  &=& (-1)^{d+i} \bar w_{i-1-n-q}^{q-1}, &\quad i=2,\ldots,r, &\quad q=1,\ldots i-1-n,\\
	\Lm \bar w_{i-1-n-q}^{-q-1} &=& - \bar w _{i-1-n-q}^{-q-2}, &\quad i=1,\ldots,r-1, &\quad q=0,\ldots.i-1-n, \\ 
	\Lm \bar w_{r-1-n-q}^{-q-1} &=& 0, &&\quad q=0,\ldots,r-1-n.
\ea\ee
Due to $[\Lm,s]=0$ the corresponding multiplier fields transform as 
\bea
	\Lm \Pi^\alpha=-s\Lm \bar C^\alpha,
\eea 
for some element of the  BRST-doublets $(\bar C^\alpha,\Pi^\alpha)$. Finally, the \L-variations of the external sources $\rho^*_a$ are
\be\ba{rclll}
	\Lm \tau_{1+q}^{*-q-1} &=& 0, &&\quad q=0,\ldots,d-2,\\
	\Lm \tau_{i+q}^{*-q-1} &=& (-1)^i \tau_{i+q}^{*-q-2}, &\quad i=2,\ldots,r, &\quad q=0,\ldots,d-i-1,\\
	\Lm \eta_{d-i+1+q}^{*-q-1} &=& \eta_{d-i+1+q}^{*-q-2}, &\quad i=1,\ldots,r-1, &\quad q=0,\ldots,i-2,\\
	\Lm \eta_{d-r+1+q}^{*-q-1} &=& 0, &&\quad q=0,\ldots,r-2.
\label{L4}
\ea\ee
The algebra $[\Lm,s]=0$ on the classical fields is valid on-shell
\bea
	\left[\Lm,s\right] X_{d-i} = (-1)^{d(i+1)}\W{\G{0}}{Y_i}, \qquad \left[\Lm,s\right] Y_{i-1} = (-1)^{d+i}\W{\G{0}}{X_{d-i+1}}.
\eea
The inclusion of the external sources also yields a linear breaking of the Ward-identity (\cf(\ref{WI})) associated to the latter presented \L-transformations (\ref{L1})--(\ref{L4})
\bea
	{\cal W}^{\Lm}\G{0}=\Delta^{\Lm},
\eea
where
\bea
	\Delta^{\Lm} &=& \intM{d}\left\{\sum_{i=1}^{r-1}(-1)^{i+1}\kappa_{d-i-1}d*\eta_{d-i}^{*-1}+ \sum_{i=2}^r (-1)^{d(i+1)}\lambda_{i-2}d*\tau_{i-1}^{*-1}\right\}. 
\eea
The symmetry property of the above transformations can also be understood in another context. The action (\ref{actionredef}) can be seen as the dimensional reduction of a similar model in $d+1$ space-time dimensions. In this sense the \L-symmetry is nothing else but the leftover of the surplus \VSUSY in the reduced direction. Hence, it is obvious that \L~is indeed a symmetry transformation. 

\section{Conclusion}

In this short note we applied the procedure of \cite{Wallet:Ikemori:Dayi} to the case of an arbitrary topological $p$-form model of Schwarz-type in $d$ space-time dimensions \cite{Baulieu}. We presented the \BRSTs in terms of generic form fields, and we extended the formalism to the derivation of the \VSUSYs and a scalar supersymmetry called \L-symmetry.\newline
Obviously, the model under consideration incorporates an ordinary BF theory in arbitrary dimensions \cite{Lucchesi:PiguetSorella,PiguetSorella} by setting all fields except $B_p \equiv X_{d-2}$ and $A \equiv Y_1$ to zero. As a three-dimensional application of the generic $p$-form model one can reconstruct the results for the so-called BFK model \cite{DelCima:1999kb}. With a similar method the authors of \cite{Pisar:2000xs} analyzed the BFK model enlarged by a Chern-Simons term. 

\section{Acknowledgements}

It is a great pleasure to thank Jesper Grimstrup for discussions and comments.


\providecommand{\href}[2]{#2}\begingroup\raggedright\endgroup

\end{document}